\documentclass[12pt]{article}
\usepackage{amssymb,amsmath,epsfig}
\allowdisplaybreaks

\begin{document}

\title{\bf Thermodynamics of Various Entropies in Specific Modified Gravity with Particle Creation}
\author{\textbf{Abdul Jawad} \thanks {jawadab181@yahoo.com; abduljawad@ciitlahore.edu.pk},
\textbf{Shamaila Rani} \thanks {shamailatoor.math@yahoo.com} and
\textbf{Salman Rafique} \thanks{salmanmath004@gmail.com}\\
Department of Mathematics, COMSATS Institute of\\ Information
Technology, Lahore-54000, Pakistan.}

\date{}
\maketitle
\begin{abstract}

We consider the particle creation scenario in the dynamical
Chern-Simons modified gravity in the presence of perfect fluid
equation of state $p=(\gamma-1)\rho$. By assuming various modified
entropies (Bekenstein, logarithmic, power law correction and Reyni),
we investigate the first law of thermodynamics and generalized
second law of thermodynamics on the apparent horizon. In the
presence of particle creation rate, we discuss the generalized
second law of thermodynamics and thermal equilibrium condition. It
is found that thermodynamic laws and equilibrium condition remain
valid under certain conditions of parameters.

\end{abstract}

\section{Introduction}

Recently, different observations such as cosmic microwave background
radiations (CMBR) \cite{1} and sloan digital sky survey (SDSS)
\cite{2} have confirmed the accelerated expansion of universe. It is
predicted that the expansion of the universe is due the curious form
of force, i.e, dark energy (DE). This discovery was unexpected,
because before this invention cosmologists just think that the
expansion of the universe would be decelerating because of the
gravitational attraction of the matter in the universe. Scientists
have proposed various aspects of dynamical DE models such as
quintessence \cite{3}, K-essence \cite{4}, phantom \cite{5}, quintom
\cite{6}, tachyon \cite{7}, holographic \cite{8} and pilgrim DE
\cite{9}-\cite{S13}. The simplest candidate of DE is cosmological
constant but its formation and mechanisms are unknown. In order to
explain the cosmic acceleration, various DE models and modified
theories of gravity have been predicted such as $f(R)$, $f(T)$
\cite{S14}-\cite{S16}, $f(T,T_G$) \cite{S17,S18}, $f(T,\textrm{T})$
\cite{S19,S20}, dynamical Chern-Simons modified gravity
\cite{S46}-\cite{S49} etc.

In cosmology, attempts to disclose the connection between Einstein
gravity and thermodynamics were carried out in
\cite{S20aa}-\cite{S20l}. The basic concept of thermodynamics comes
from black hole physics. In general, there has been some profound
thought on the connection among gravity and thermodynamics for a
long time. The initial work was done by Jacobson who showed that the
gravitational Einstein equation can be derived from the relation
between the horizon area and entropy, together with the Clausius
relation $\delta Q=T\delta S$ (where $\delta Q$, $T$ and $\delta S$
represents the change in energy, temperature and entropy change o
the system respectively) \cite{S22}. Further, various gravity
theories has been investigated to study the deep connection between
gravity and thermodynamics \cite{S21a}-\cite{S21h}. Cosmological
investigations of thermodynamics in modified gravity theories have
been executed in Refs.~\cite{CT}-\cite{CT6} (for a recent review on
thermodynamic properties of modified gravity theories, see,
e.g.,~\cite{Bamba:2016aoo}). Saha and Mondal \cite{r1} have studied
thermodynamics on apparent horizon with the help of gravitationally
induced particle scenario with constant specific entropy and
particle creation rate $\Gamma$.

In the present work, we develop the particle creation scenario in
the dynamical Chern-Simons modified gravity by assuming various
modified entropies (bekenstein entropy logarithmic entropy, power
law correction and Reyni entropy) on the apparent horizon. In the
presence of particle creation rate, we discuss the generalized
second law of thermodynamics (GSLT). This paper is organized as
follows: In the next section, we will present the basic equations of
dynamical Chern-Simons modified theory and particle creation rate.
Also, we investigate the first law of thermodynamics. Section
\textbf{3} contains the illustration of GSLT. In section \textbf{4},
we analyze the stability of thermodynamical equilibrium for constant
as well as variable $\Gamma$. Section \textbf{5} contains the
comparison of results with preceding works. In the last section, we
summarize our results.

\section{Modified Entropies and Particle Creation Rate}

It was shown that the differential form of the Friedmann equation in
the FRW universe can be written in the form of the first law of
thermodynamics on the apparent horizon. The profound connection
provides a thermodynamical interpretation of gravity which makes it
interesting to explore the cosmological properties through
thermodynamics. It was proved that for any spherically symmetric
spacetime, the field equations can be expressed as $TdS=dE+PdV$ for
any horizon \cite{S23}, where $E$, $P$ and $V$ represent the
internal energy, pressure and volume of the spherical system
respectively. The generalized second law of thermodynamics (GSLT)
has been studied extensively in the behavior of expanding universe.
According to GSLT, '' \textit{the sum of all entropies of the
constituents (mainly DM and DE) and entropy of boundary (either it
is apparent or event horizons) of the universe can never
decrease.}'' \cite{S24}. Most of the researchers have discussed the
validity of GSLT of different systems including interaction of two
fluid components dark energy (DE) and dark matter (DM) \cite{S28}
and interaction of three fluid components \cite{S29} in FRW
universe.

To discuss the behavior of GSLT, scientists assumed the horizon
entropy as $1/4$ of its area \cite{S25}, power law correction
\cite{S26}, logarithmic entropy \cite{S27} and Reyni entropy. GSLT
has been discussed on the basis of gravitationally induced particle
scenario which was firstly introduced by Schrodinger \cite{S30} on
microscopic level. Parker at al. \cite{S31}-\cite{S35} extend this
mechanism towards quantum field theory in curved space. Prigogine at
al. \cite{S36} introduced the macroscopic mechanism of
gravitationally induced particle scenario. Afterward, covariant
description and difference between particle creation and bulk
viscosity of creation process was given \cite{S37}-\cite{S39}. The
particle creation process can be predicted with the incorporation of
backreaction term in the Einstein field equations whose negative
weight may help in clarifying the cosmic acceleration. In such a
way, most of the phenomenological models of particle creation have
been granted \cite{S40}-\cite{S16a}. In addition, it was proved that
phenomenological particle creation help us to discuss the behavior
of accelerating universe and paved the alternative way to the
concordance $\Lambda$CDM model \cite{S16a}-\cite{S20a}.

In the following discussion, we check the validity of first law of
thermodynamics with Gibbs relation, GSLT and thermodynamical
equilibrium by assuming the following entropy corrections.
\begin{itemize}
\item \textbf{Bekenstein entropy:}
The Bekenstein entropy and Hawking temperature of the apparent
horizon are given by $(8\pi=G=1)$
\begin{eqnarray}\label{E11}
S_{\textmd{A}}=\frac{A}{4}
=\frac{R^2_A}{8}~~~~\text{and}~~~~T_{\textmd{A}}=\frac{1}{2\pi R_A}
=\frac{4}{R_A} ~~~~\text{where}~~~~A=4\pi R^2_A.
\end{eqnarray}
\item \textbf{Logarithmic corrected entropy:}
To study the expansion of entropy of the universe, we discuss the
addition of entropy related to the horizon. Quantum gravity allows
the logarithmic corrections in the presence of thermal equilibrium
fluctuations and quantum fluctuations \cite{15}-\cite{21}. The
logarithmic entropy corrections can be defined as
\begin{eqnarray}\label{E22}
S_A=\frac{A}{4L^2_p}+\alpha\ln\frac{A}{4L^2_p}+\beta\frac{4L^2_p}{A},
\end{eqnarray}
where $L_P$ is the Planck's length and $\alpha,~\beta$ are constants
whose values are still under consideration.
\item \textbf{Power law corrected entropy:}
Thermodynamics of apparent horizon in the standard FRW cosmology,
the geometric entropy is assumed to be proportional to its horizon
area $(S_A=\frac{A}{4})$. The quantum corrections provided to the
entropy-area relationship lead to the curvature correction in the
Einstein-Hilbert action and vice versa \cite{22}. The power-law
quantum correction to the horizon entropy motivated by the
entanglement of quantum fields between inside and outside of the
horizon is given by \cite{23}
\begin{eqnarray}\label{E26}
S_{\textmd{A}}=\frac{A}{4L^2_{\textmd{p}}}\big(1-K_\delta
A^{1-\frac{\delta}{2}}\big),
\end{eqnarray}
where
$K_\delta=\frac{\delta\big(4\pi\big)^{\frac{\delta}{2}-1}}{\big(4-\delta\big)r^{4-\delta}_c}$.
Here $ \delta $ is a dimensionless constant and $r_c$ is the
crossover scale.
\item \textbf{Renyi entropy:}
A novel sort of Renyi entropy has been proposed and inspected
\cite{24}-\cite{26}. In which not exclusively is the logarithmic
corrected entropy of the original Renyi entropy utilized yet the
Bekenstein– Hawking entropy $S_{BH}$ is thought to be a
non-extensive Tsallis entropy $S_A$. One can obtain Renyi entropy
$S_R$ as \cite{26}
\begin{eqnarray}\label{E34}
S_R=\frac{\ln(1+\eta S_A)}{\eta}.
\end{eqnarray}
\end{itemize}

The action of Chern-Simons theory is given by \cite{S46}-\cite{S48}
\begin{eqnarray}\label{C1}
S=\frac{1}{16\pi G}\int
d^4x\bigg[\sqrt{-g}R+\frac{\ell}{4}\theta^\star
R^{\rho\sigma\mu\nu}R_{\rho\sigma\mu\nu}-\frac{1}{2}g^{\mu\nu}\nabla_\mu\theta\nabla\nu\theta+V(\theta)\bigg]
+S_{mat},
\end{eqnarray}
where $R$, $^\star R^{\rho\sigma\mu\nu}R_{\rho\sigma\mu\nu}$,
$\ell$, $\theta$, $S_{mat}$, and $V(\theta)$ are Ricci scalar, a
topological invariant called the Pontryagin term, the coupling
constant, the dynamical variable, the action of matter and the
potential, respectively. The Freidmann equation for flat universe
turns out to be \cite{S49}
\begin{eqnarray}\label{E1}
H^2=\frac{1}{3}\rho+\frac{c^2}{6a^6},
\end{eqnarray}
here $c$ is constant, $H=\frac{\dot{a}(t)}{a(t)}$ is the Hubble
parameter and $a(t)$ is the scale factor. The equation of continuity
for this model can be described as
\begin{eqnarray}\label{E3}
\dot{\rho}+\Theta\big(\rho+P+\Pi\big)=0.
\end{eqnarray}
The particle creation pressure $(\Pi)$ representing the
gravitationally induced process for particle creation and
$\Theta=3H$ is the fluid expansion. The total number of
$n-$particles in an open thermodynamics are
\begin{eqnarray}\label{E4}
\dot{n}+\Theta n=n\Gamma,
\end{eqnarray}
where $\Gamma$ is the creation rate of number of particles in
comoving volume i.e., $(N=na^3)$ having two phases negative and
positive. The negative $\Gamma$ relates with the particle anhilation
and the positive $\Gamma$ relates with production of particle.
Equations (\ref{E3}) and (\ref{E4}) with Gibbs relations can be
written as
\begin{eqnarray}\label{E5}
Tds=d\bigg(\frac{\rho}{n}\bigg)+pd\bigg(\frac{1}{n}\bigg).
\end{eqnarray}
The equation related to creation pressure $\Pi$ and $\Gamma$ can be
determined as
\begin{eqnarray}\label{E6}
\Pi=-\frac{\Gamma}{\Theta}\big(\rho+p\big).
\end{eqnarray}
Under traditional assumption that the specific entropy of each
particle is constant, i.e., the process is adiabatic or isentropic,
which implies that dissipative fluid is similar to a perfect fluid
with a non-conserved particle number. The respective EoS for this
model represented by $p=(\gamma-1)\rho$. Differentiation of
Eq.(\ref{E1}) gives
\begin{eqnarray}\label{E2}
\dot{H}=-\frac{1}{2}\bigg((\rho+p+\Pi)-H\frac{c^2}{a^6}\bigg).
\end{eqnarray}
Inserting Eqs.(\ref{E6}) and $p=(\gamma-1)\rho$ in above equation,
we get
\begin{eqnarray}\label{E7}
\frac{\dot{H}}{H^2}=-\frac{1}{2H^2}\bigg(\gamma\big(3H^2-\frac{c^2}{2a^6}\big)
\big(1-\frac{\Gamma}{3H}\big)+\frac{c^2}{a^6}\bigg).
\end{eqnarray}
For flat FRW universe, Hubble parameter relates with the apparent
horizon as $R_{\textmd{A}}=\frac{1}{H}$. Differentiating the
apparent horizon with respect to time, we have
\begin{eqnarray}\label{E10}
\dot{R_A}=-\frac{\dot{H}}{H^2}
=\frac{1}{2H^2}\bigg(\gamma\big(3H^2-\frac{c^2}{2a^6}\big)
\big(1-\frac{\Gamma}{3H}\big)+\frac{c^2}{a^6}\bigg).
\end{eqnarray}
The deceleration parameter $q$ is of the form
\begin{eqnarray}\label{E8}
q=-\frac{\dot{H}}{H^2}-1=\frac{1}{2H^2}\bigg(\gamma\big(3H^2-\frac{c^2}{2a^6}\big)
\big(1-\frac{\Gamma}{3H}\big)+\frac{c^2}{a^6}\bigg)-1.
\end{eqnarray}

\subsection{First Law of Thermodynamics}

Next, we investigate the first law of thermodynamics in the presence
of modified entropies. The relation between thermodynamics and
Einstein field equations was found by Jacobson with the help of
clausius relation at apparent horizon described as
\begin{eqnarray}\label{E12}
-dE_{\textmd{A}}=T_{\textmd{A}}dS_{\textmd{A}}.
\end{eqnarray}
For the sake of convenance we consider
$X=T_{\textmd{A}}dS_{\textmd{A}}+dE_{\textmd{A}}$. The differential
$dE_{\textmd{A}}$ is the amount of energy crossing the apparent
horizon can be evaluated as \cite{13}
\begin{eqnarray}\label{E13}
-dE_{\textmd{A}}=\frac{1}{2}R^3(\rho+p)Hdt=\frac{\gamma}{2H^2}\big(3H^2-\frac{c^2}{2a^6}\big).
\end{eqnarray}\\
\textbf{\underline{Bekenstein entropy:}}\\\\ From Eq.(\ref{E11}),
the differential of surface entropy at apparent horizon leads to
\begin{eqnarray}\label{E14}
dS_A=\frac{1}{8H^3}\bigg(\gamma\big(3H^2-\frac{c^2}{2a^6}\big)
\big(1-\frac{\Gamma}{3H}\big)+\frac{c^2}{a^6}\bigg).
\end{eqnarray}
The above equation with horizon temperature leads to
\begin{eqnarray}\label{E15}
T_A dS_A= \frac{1}{2H^2}\bigg(\gamma\big(3H^2-\frac{c^2}{2a^6}\big)
\big(1-\frac{\Gamma}{3H}\big)+\frac{c^2}{a^6}\bigg).
\end{eqnarray}
Hence $X$ becomes
\begin{eqnarray}\label{E16}
X=-\frac{\gamma}{2H^2}\big(3H^2-\frac{c^2}{2a^6}\big)+\frac{1}{2H^2}\bigg(\gamma\big(3H^2-\frac{c^2}{2a^6}\big)
\big(1-\frac{\Gamma}{3H}\big)+\frac{c^2}{a^6}\bigg).
\end{eqnarray}
With the help of Eq.(\ref{E16}), we observe that the first law of
thermodynamics holds (i.e., $X\rightarrow0$) when
\begin{eqnarray}\label{E16+}
\Gamma=3H\bigg(\frac{c^2}{\gamma
a^6}\big(3H^2-\frac{c^2}{2a^6}\big)^{-1}\bigg).
\end{eqnarray}
\textbf{\underline{Logarithmic corrected entropy:}}\\\\
The differential form of Eq.(\ref{E22}) is given as
\begin{eqnarray}\label{E23}
dS_A=\frac{\bigg(\gamma\big(3H^2-\frac{c^2}{2a^6}\big)
\big(1-\frac{\Gamma}{3H}\big)+\frac{c^2}{a^6}\bigg)\bigg(\frac{1}{4HL^2_p}+2\alpha
H-16\beta H^3L^2_p\bigg)}{2H^2}dt,
\end{eqnarray}
which leads to
\begin{eqnarray}\label{E24}
T_AdS_A=\frac{1}{2H^2}\bigg(\gamma\big(3H^2-\frac{c^2}{2a^6}\big)
\big(1-\frac{\Gamma}{3H}\big)+\frac{c^2}{a^6}\bigg)\bigg(\frac{1}{L^2_{\textmd{p}}}+8H^2\alpha-64\beta
H^4L^2_{\textmd{p}}\bigg).
\end{eqnarray}
Combining Eqs.(\ref{E13}) and (\ref{E24}), we get
\begin{eqnarray}\label{E24a}
X&=&\frac{1}{2H^2}\bigg(\gamma\big(3H^2-\frac{c^2}{2a^6}\big)
\big(1-\frac{\Gamma}{3H}\big)+\frac{c^2}{a^6}\bigg)\bigg(\frac{1}{L^2_{\textmd{p}}}+8H^2\alpha-64\beta
H^4L^2_{\textmd{p}}\bigg)\nonumber\\
&-&\frac{\gamma}{2H^2}\big(3H^2-\frac{c^2}{2a^6}\big).
\end{eqnarray}
From above equation, we observe the validity of first  law of
thermodynamics if
\begin{eqnarray}\label{E24a+}
\Gamma=3H\bigg(1-\frac{1}{\gamma}\big(3H^2-\frac{c^2}{a^6}\big)^{-1}
\big(\frac{\gamma(3H^2-\frac{c^2}{2a^6})}{\frac{1}{L^2_{\textmd{p}}}+8H^2\alpha-64\beta
H^4L^2_{\textmd{p}}}-\frac{c^2}{a^6}\big)\bigg).
\end{eqnarray}
\textbf{\underline{Power law corrected entropy:}}\\\\
Differentiating Eq.(\ref{E26}), we get
\begin{eqnarray}\label{E27}
dS_A=\frac{1}{2H^2}\bigg(\gamma\big(3H^2-\frac{c^2}{2a^6}\big)
\big(1-\frac{\Gamma}{3H}\big)+\frac{c^2}{a^6}\bigg)
\bigg(\frac{1}{4HL^2_{\textmd{p}}}-\frac{K_\delta}{4L^2_{\textmd{p}}}
\big(2-\frac{\delta}{2}\big)\big(\frac{1}{H}\big)^{3-\delta}\bigg)dt,
\end{eqnarray}
which yields
\begin{eqnarray}\label{E28}
T_AdS_A=\frac{1}{2H^2}\bigg(\gamma\big(3H^2-\frac{c^2}{2a^6}\big)
\big(1-\frac{\Gamma}{3H}\big)+\frac{c^2}{a^6}\bigg)
\bigg(\frac{1}{L^2_{\textmd{p}}}-\big(2-\frac{\delta}{2}\big)\frac{K_\delta}{L^2_p}
\big(\frac{1}{H}\big)^{2-\delta}\bigg).
\end{eqnarray}
Hence, $X$ takes the form
\begin{eqnarray}\label{E28a}
X&=&\frac{1}{2H^2}\bigg(\gamma\big(3H^2-\frac{c^2}{2a^6}\big)
\big(1-\frac{\Gamma}{3H}\big)+\frac{c^2}{a^6}\bigg)
\bigg(\frac{1}{L^2_{\textmd{p}}}-\big(2-\frac{\delta}{2}\big)\frac{K_\delta}{L^2_p}
\big(\frac{1}{H}\big)^{2-\delta}\bigg)\nonumber\\
&-&\frac{\gamma}{2H^2}\big(3H^2-\frac{c^2}{2a^6}\big).
\end{eqnarray}
From Eq.(\ref{E28a}), it can be analyzed that the first law of
thermodynamics remains valid for the following particle creation
rate
\begin{eqnarray}\nonumber
\Gamma&=&3H\bigg(1-\frac{1}{\gamma}\big(3H^2-\frac{c^2}{2a^6}\big)^{-1}\bigg(\gamma\big(3H^2-\frac{c^2}{2a^6}\big)
\bigg(\frac{1}{L^2_{\textmd{p}}}-(2-\frac{\delta}{2})\frac{K_\delta}{L^2_p}
(\frac{1}{H})^{2-\delta}\bigg)^{-1}\nonumber\\
&-&\frac{c^2}{a^6}\bigg)\bigg).
\end{eqnarray}\\\\
\textbf{\underline{Renyi entropy:}}\\\\
Eq.(\ref{E34}) gives the differential of entropy as
\begin{eqnarray}\label{E35}
dS_R=\frac{1}{H}\bigg(\gamma\big(3H^2-\frac{c^2}{2a^6}\big)
\big(1-\frac{\Gamma}{3H}\big)+\frac{c^2}{a^6}\bigg)\bigg(\frac{1}{\eta+8H^2}\bigg),
\end{eqnarray}
which gives rise to
\begin{eqnarray}\label{E36}
T_AdS_R=\frac{4}{(\eta+8H^2)}\bigg(\gamma\big(3H^2-\frac{c^2}{2a^6}\big)\big(1-\frac{\Gamma}{3H}\big)+
\frac{c^2}{a^6}\bigg).
\end{eqnarray}
Using Eqs.(\ref{E13}) and (\ref{E36}) we get
\begin{eqnarray}\label{E37}
X=\frac{4}{(\eta+8H^2)}\bigg(\gamma\big(3H^2-\frac{c^2}{2a^6}\big)\big(1-\frac{\Gamma}{3H}\big)+
\frac{c^2}{a^6}\bigg)-\frac{\gamma}{2H^2}\big(3H^2-\frac{c^2}{2a^6}\big).
\end{eqnarray}
From Eq.(\ref{E37}), it can be seen that the first law of
thermodynamics is showing the validity when
\begin{eqnarray}\label{E37+}
\Gamma=3H\bigg(1-\frac{1}{\gamma}\big(3H^2-\frac{c^2}{2a^6}\big)^{-1}\big(\frac{\gamma(\eta+8H^2)}{8H^2}
(3H^2-\frac{c^2}{2a^6})-\frac{c^2}{a^6}\big)\bigg).
\end{eqnarray}

\section{Generalized Second Law of Thermodynamics}

We discuss the GSLT of an isolated macroscopic physical system where
the total entropy $S_T$ must satisfies the following conditions
$d(S_{\textmd{A}}+S_{\textmd{f}})\geq0$ i.e., entropy function
cannot be decrease. In this relation, $S_{\textmd{A}}$ and
$S_{\textmd{f}}$ appear as the entropy at apparent horizon and the
entropy of cosmic fluid enclosed within the horizon, respectively.
The Gibbs equation is of the form
\begin{eqnarray}\label{E17}
T_{\textmd{f}}dS_{\textmd{f}}=dE_{\textmd{f}}+p dV,
\end{eqnarray}
where $ T_{\textmd{f}} $ is the temperature of the cosmic fluid and
$ E_{\textmd{f}}$ is the energy of the fluid ($E_{\textmd{f}}=\rho
V) $. The evolution equation for fluid temperature having constant
entropy can be described as \cite{14}
\begin{eqnarray}\label{E18}
\frac{\dot{T_{\textmd{f}}}}{T_{\textmd{f}}}=(\Gamma-\Theta)\frac{\partial
p}{\partial \rho}.
\end{eqnarray}
Using Eq.(\ref{E7}), we get
$(\Gamma-\Theta)=\frac{6H\dot{H}+\frac{3Hc^2}{a^6}}{\gamma\big(3H^2-\frac{c^2}{2a^6}\big)}$,
hence, the above Eq.(\ref{E18}) leads to the integral
\begin{eqnarray}\label{E18a}
\ln\bigg(\frac{T_{\textmd{f}}}{T_0}\bigg)=\frac{\gamma-1}{\gamma}\int\frac{\big(2H\dot{H}
+\frac{Hc^2}{a^6}\big)}{\big(H^2-\frac{c^2}{6a^6}\big)}dH.
\end{eqnarray}
Integration of above equation leads to
\begin{eqnarray}\label{E19}
T_{\textmd{f}}=T_0\bigg(H^2-\frac{c^2}{6a^6}\bigg)^{\frac{\gamma-1}{\gamma}},
\end{eqnarray}
where $T_0$ is the constant of integration. Equation (\ref{E17})
yields the differential of fluid entropy as
\begin{eqnarray}\label{E20}
dS_f=\frac{T^{-1}_0}{H^2}\bigg(H^2-\frac{c^2}{6a^6}\bigg)^{\frac{1-\gamma}{\gamma}}
\bigg(\big(2\dot{H}+\frac{c^2}{a^6}\big)-\frac{\gamma\dot{H}}{H^2}\big(3H^2-\frac{c^2}{2a^6}\big)\bigg).
\end{eqnarray}
Next, we observe the validity GSLT by assuming Bekenstein entropy,
logarithmic corrected entropy, power law correction and Renyi
entropy.\\\\
\textbf{\underline{Bekenstein entropy:}}\\\\
In present case, we get the differential of total entropy by using
Eqs.(\ref{E14}) and (\ref{E20}) as
\begin{eqnarray}\label{E21}
\dot{S_T}&=&\frac{1}{8H^3}\bigg(\gamma\big(3H^2-\frac{c^2}{2a^6}\big)
\big(1-\frac{\Gamma}{3H}\big)+\frac{c^2}{a^6}\bigg)+\frac{T^{-1}_0}{H^2}
\bigg(H^2-\frac{c^2}{6a^6}\bigg)^{\frac{1-\gamma}{\gamma}}\nonumber\\
&\times&\bigg(\big(2\dot{H}+\frac{c^2}{a^6}\big)-\frac{\gamma\dot{H}}{H^2}
\big(3H^2-\frac{c^2}{2a^6}\big)\bigg),
\end{eqnarray}
where $S_{T}=S_{\textmd{A}}+S_{\textmd{f}}$.
\begin{figure} \centering
\epsfig{file=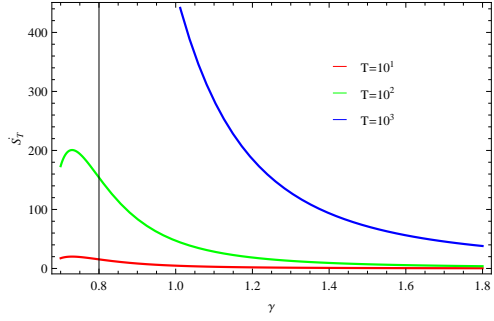,width=.50\linewidth}\caption{Plot of
$\dot{S_T}$ versus $\gamma$ for Bekenstein entropy.}
\end{figure}
The plot between $\dot{S_T}$ versus $\gamma$ is shown in Figure
\textbf{1} for three values of $T$. We observe the validity of GSLT
at the present epoch by setting the constant values $H=H_0=67$,
$c=-1$, $a=a_0=1$ and $q=-0.53$. It can be analyzed from figure that
$\dot{S_T}\geq0$ for all values of $T$ leads to the validity of
GSLT.\\\\
\textbf{\underline{Logarithmic corrected entropy:}}\\\\
Using Eqs.(\ref{E23}) and (\ref{E20}) the differential of total
entropy is given by
\begin{eqnarray}\label{E25}
\dot{S_T}&=&\frac{1}{2H^2}\bigg(\gamma\big(3H^2-\frac{c^2}{2a^6}\big)
\big(1-\frac{\Gamma}{3H}\big)+\frac{c^2}{a^6}\bigg)\bigg(\frac{1}{4HL^2_p}+2\alpha
H-16\beta H^3L^2_p\bigg)dt\nonumber\\
&+&\frac{T^{-1}_0}{H^2}\bigg(H^2-\frac{c^2}{6a^6}\bigg)^{\frac{1-\gamma}{\gamma}}
\bigg(\big(2\dot{H}
+\frac{c^2}{a^6}\big)-\frac{\gamma\dot{H}}{H^2}\big(3H^2-\frac{c^2}{2a^6}\big)\bigg)dt.
\end{eqnarray}
\begin{figure} \centering
\epsfig{file=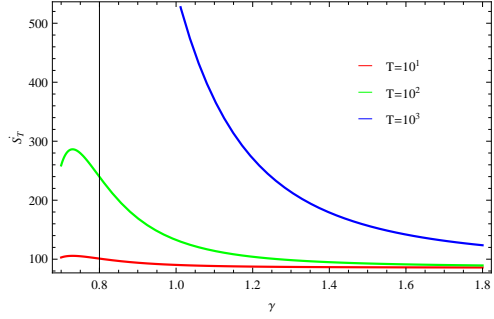,width=.50\linewidth}\caption{Plot of
$\dot{S_T}$ versus $\gamma$ for logarithmic corrected entropy.}
\end{figure}The plot of $\dot{S_T}$ versus $\gamma$ with respect to the three values
of $T$ by fixing the constant values $\alpha=1$, $\beta=-0.00001$
and $L_P=1$ as shown in Figure \textbf{2}. The trajectories in the
plot remains which leads to the validity of GSLT in the presence of
logarithmic entropy.\\\\
\textbf{\underline{Power law corrected entropy:}}\\\\
From Eqs.(\ref{E28}) and (\ref{E20}), we get the differential of
total entropy as
\begin{eqnarray}\label{E29}
\dot{S_T}&=&\frac{1}{2H^2}\bigg(\gamma\big(3H^2-\frac{c^2}{2a^6}\big)
\big(1-\frac{\Gamma}{3H}\big)+\frac{c^2}{a^6}\bigg)
\bigg(\frac{1}{4HL^2_{\textmd{p}}}-\frac{K_\delta}{4L^2_{\textmd{p}}}
\big(2-\frac{\delta}{2}\big)\big(\frac{1}{H}\big)^{3-\delta}\bigg)dt\nonumber\\
&+&\frac{T^{-1}_0}{H^2}\bigg(H^2-\frac{c^2}{6a^6}\bigg)^{\frac{1-\gamma}{\gamma}}
\bigg(\big(2\dot{H}+\frac{c^2}{a^6}\big)-\frac{\gamma\dot{H}}{H^2}\big(3H^2-\frac{c^2}{2a^6}\big)\bigg)dt.
\end{eqnarray}
\begin{figure} \centering
\epsfig{file=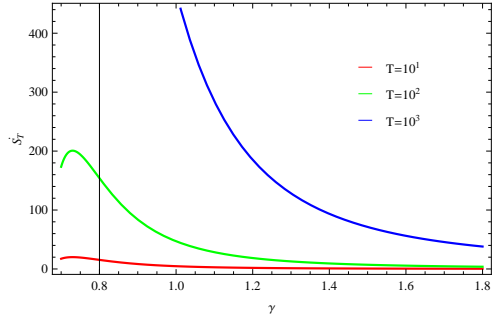,width=.50\linewidth}\caption{Plot of
$\dot{S_T}$ versus $\gamma$ for power law corrected entropy.}
\end{figure}
The plot of $\dot{S_T}$ versus $\gamma$ for three values of $T$ by
setting the constant values as $\delta=1$, $r_c=\frac{1}{67}$, and
$L_P=1$ as shown in Figure \textbf{3}. We can see $\dot{S_T}$
remains positive for all values of $T$ which leads to the validity
of GSLT.\\\\
\textbf{\underline{Renyi entropy:}}\\\\
We observe the validity of GSLT with $\dot{S}_T\geq0$ for which
Eqs.(\ref{E35}) and (\ref{E20}) takes the form
\begin{eqnarray}\label{E38}
\dot{S_T}&=&\frac{1}{
H(\eta+8H^2)}\bigg(\gamma\big(3H^2-\frac{c^2}{2a^6}\big)
\big(1-\frac{\Gamma}{3H}\big)+\frac{c^2}{a^6}\bigg)
+\frac{T^{-1}_0}{H^2}\big(H^2-\frac{c^2}{6a^6}\big)^{\frac{1-\gamma}{\gamma}}\nonumber\\
&\times&\bigg(\big(2\dot{H}+\frac{c^2}{a^6}\big)-\frac{\gamma\dot{H}}{H^2}\big(3H^2-\frac{c^2}{2a^6}\big)\bigg).
\end{eqnarray}
\begin{figure} \centering
\epsfig{file=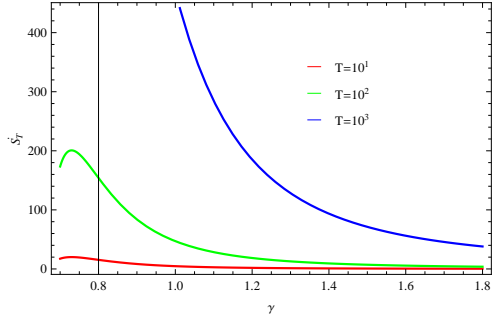,width=.50\linewidth}\caption{Plot of
$\dot{S_T}$ versus $\gamma$ Renyi entropy.}
\end{figure}The plot between $\dot{S_T}$ versus $\gamma$ for three
values of is shown in Figure \textbf{4} for $\eta=1$. It can be seen
that GSLT is satisfying $\dot{S_T}\geq0$ for all values of $T$ which
leads to the validity of GSLT.

\section{Thermodynamical equilibrium}

We discuss the two scenarios for thermodynamical equilibrium by
taking particle creation rate $\Gamma$=constant and
$\Gamma=\Gamma(t)$ for which entropy function attains a maximum
entropy state i.e., $d(S_{\textmd{A}}+S_{\textmd{f}})<0$.

\subsection{$\Gamma$ = constant}
Firstly, we consider $\Gamma$ as constant and observe the stability
of thermodynamical equilibrium.\\\\
\textbf{\underline{Bekenstein entropy:}}\\\\
We get second order differential equation for constant $\Gamma$ by
using Eq.(\ref{E21})
\begin{eqnarray}\label{E21a}
\ddot{S_T}&=&-\frac{3\dot{H}\bigg(\frac{c^2}{a^6}+\gamma\big(1-\frac{\Gamma}{3H}\big)\big(-\frac{c^2}{2a^6}+
3H^2\big)\bigg)}{8H^4}-2\dot{H}\big(-\frac{c^2}{6a^6}+H^2\big)^{\frac{1-\gamma}{\gamma}}\nonumber\\
&\times&\frac{\bigg(-\gamma\big(1-\frac{\Gamma}{3H}\big)\big(3H^2-\frac{c^2}{2a^6}\big)
+\frac{\gamma\big(3H^2-\frac{c^2}{2a^6}\big)
\big(\frac{c^2}{a^6}+\gamma\big(1-\frac{\Gamma}{3H}\big)\big(3H^2-\frac{c^2}{2a^6}\big)\big)}{2H^2}\bigg)}
{H^3T_0}\nonumber\\
&+&\frac{\bigg(-\gamma\big(1-\frac{\Gamma}{3H}\big)
\big(3H^2-\frac{c^2}{2a^6}\big)+\frac{\gamma\big(3H^2-\frac{c^2}{2a^6}\big)
\big(\frac{c^2}{a^6}+\gamma\big(1-\frac{\Gamma}{3H}\big)\big(3H^2-\frac{c^2}{2a^6}\big)\big)}{2H^2}\bigg)}{\gamma
H^2T_0}\nonumber\\
&\times&\big(\frac{c^2\dot{a}}{a^7}+2H\dot{H}\big)(1-\gamma)\big(H^2-\frac{c^2}{6a^6}\big)^{-1+\frac{1-\gamma}{\gamma}}
+\gamma\Gamma\dot{H}\big(3H^2-\frac{c^2}{2a^6}\big)\frac{1}{24H^5}\nonumber\\
&-&\frac{6c^2\dot{a}}{8H^3a^7}+\frac{\gamma\big(1-\frac{\Gamma}{3H}\big)\big(\frac{3c^2\dot{a}}{a^7}+6H\dot{H}\big)}{8H^3}
+\frac{\big(H^2-\frac{c^2}{6a^6}\big)^{\frac{1-\gamma}{\gamma}}}{H^2T_0}\bigg(-\frac{\gamma\Gamma\dot{H}
}{3H^2}\nonumber\\
&\times&\big(3H^2-\frac{c^2}{2a^6}\big)-\frac{\gamma\dot{H}\big(3H^2-\frac{c^2}{2a^6}\big)\big(\frac{c^2}{a^6}+
\gamma\big(1-\frac{\Gamma}{3H}\big)
\big(3H^2-\frac{c^2}{2a^6}\big)\big)}{H^3}\nonumber\\
&-&\gamma\big(1-\frac{\Gamma}{3H}\big)\big(\frac{3c^2\dot{a}}{a^7}+6H\dot{H}\big)+\frac{\gamma\big(\frac{c^2}{a^6}+
\gamma\big(1-\frac{\Gamma}{3H}\big)\big(3H^2-\frac{c^2}{2a^6}\big)
\big)}{2H^2}\nonumber\\
&\times&\big(\frac{3c^2\dot{a}}{a^7}+6H\dot{H}\big)+\bigg(\frac{-\frac{6c^2\dot{a}}{a^7}+
\frac{\gamma\Gamma\dot{H}\big(3H^2-\frac{c^2}{2a^6}\big)}{3H^2}}{2H^2}+
\frac{\big(\frac{3c^2\dot{a}}{a^7}+6H\dot{H}\big)}{2H^2}\nonumber\\
&\times&\gamma\big(1-\frac{\Gamma}{3H}\big)\bigg)\big(3H^2-\frac{c^2}{2a^6}\big)\gamma\bigg).
\end{eqnarray}
\begin{figure} \centering
\epsfig{file=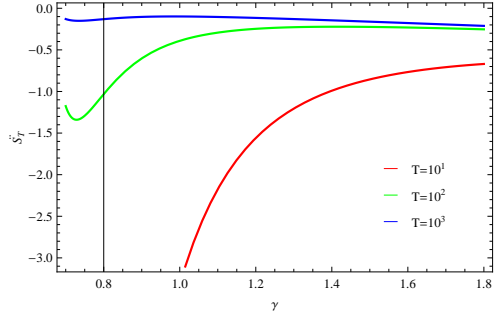,width=.50\linewidth}\caption{Plot of
$\ddot{S_T}$ versus $\gamma$ for Bekenstein entropy when $\Gamma$ is
constant.}
\end{figure}
The graphical behavior between $\ddot{S_T}$ versus $\gamma$ as shown
in Figure \textbf{5} for different values of parameter $T$. We
observe that $\ddot{S_T}<0$ for all values of $T$ which leads to the
validity of thermodynamical equilibrium.\\\\
\textbf{\underline{Logarithmic corrected entropy:}}\\\\
In the presence of logarithmic corrected entropy second order
differential equation is obtained from Eq.(\ref{E25}) as
\begin{eqnarray}\label{E25a}
\ddot{S_T}&=&-\frac{\dot{H}\big(\frac{1}{4HL^2_p}+2\alpha
H-16L^2_p\beta H^3\big)\big(\frac{c^2}{a^6}+\gamma\big(1-\frac{\Gamma}{3H}\big)\big(3H^2-\frac{c^2}{2a^6}\big)\big)}{H^3}\nonumber\\
&-&\bigg(\frac{-\gamma\big(1-\frac{\Gamma}{3H}\big)
\big(3H^2-\frac{c^2}{2a^6}\big)+\frac{\gamma\big(3H^2-\frac{c^2}{2a^6}\big)\big(\frac{c^2}{a^6}
+\gamma\big(1-\frac{\Gamma}{3H}\big)\big(3H^2-\frac{c^2}{2a^6}\big)\big)}{2H^2}}{H^3T_0}\bigg)\nonumber\\
&\times&2\dot{H}\big(H^2-\frac{c^2}{6a^6}\big)^{\frac{1-\gamma}{\gamma}}+(1-\gamma)\big(\frac{c^2\dot{a}}{a^7}+2H\dot{H}\big)
\big(H^2-\frac{c^2}{6a^6}\big)
^{-1+\frac{1-\gamma}{\gamma}}\nonumber\\
&\times&\bigg(\frac{-\gamma\big(1-\frac{\Gamma}{3H}\big)
\big(3H^2-\frac{c^2}{2a^6}\big)+\frac{\gamma\big(3H^2-\frac{c^2}{2a^6}\big)\big(\frac{c^2}{a^6}
+\gamma\big(1-\frac{\Gamma}{3H}\big)\big(3H^2-\frac{c^2}{2a^6}\big)\big)}{2H^2}}{\gamma
H^2T_0}\bigg)\nonumber\\
&+&\frac{\big(\frac{c^2}{a^6}+\gamma\big(1-\frac{\Gamma}{3H}\big)\big(3H^2-\frac{c^2}{2a^6}\big)\big)
\big(2\alpha\dot{H}-\frac{\dot{H}}{4L^2_p H^2}-48L^2_p\beta
H^2\dot{H}\big)}{2H^2}\nonumber\\
&+&\big(\frac{1}{4HL^2_p}+2\alpha H-16L^2_p\beta
H^3\big)\bigg(\frac{-\frac{6c^2\dot{a}}{a^7}+
\gamma\big(1-\frac{\Gamma}{3H}\big)\big(\frac{3c^2\dot{a}}{a^7}+6H\dot{H}\big)}{2H^2}\nonumber\\
&+&\frac{\gamma\Gamma\dot{H}\big(3H^2-\frac{c^2}{2a^6}\big)}{6H^4}\bigg)+\frac{1}{H^2T_0}\big(H^2-\frac{c^2}{6a^6}\big)
^{\frac{1-\gamma}{\gamma}}
\bigg(-\frac{\gamma\Gamma\dot{H}\big(3H^2-\frac{c^2}{2a^6}\big)}{3H^2}\nonumber\\
&-&\frac{\gamma\dot{H}\big(3H^2-\frac{c^2}{2a^6}\big)\big(\frac{c^2}{a^6}+\gamma\big(1-\frac{\Gamma}{3H}\big)
\big(3H^2-\frac{c^2}{2a^6}\big)\big)}{H^3}-\gamma\big(1-\frac{\Gamma}{3H}\big)\nonumber\\
&\times&\big(\frac{3c^2\dot{a}}{a^7}+6H\dot{H}\big)+\frac{\gamma\big(\frac{3c^2\dot{a}}{a^7}+6H\dot{H}\big)
\big(\frac{c^2}{2a^6}+\gamma\big(1-\frac{\Gamma}{3H}\big)\big(3H^2-\frac{c^2}{2a^6}\big)\big)}{2H^2}\nonumber\\
&+&\gamma\big(3H^2-\frac{c^2}{2a^6}\big)\bigg(\frac{\frac{\gamma\Gamma\dot{H}
\big(3H^2-\frac{c^2}{2a^6}\big)}{3H^2}+\gamma\big(1-\frac{\Gamma}{3H}\big)\big(\frac{3c^2\dot{a}}{a^7}+6H\dot{H}\big)}{2H^2}
\nonumber\\
&-&\frac{6c^2\dot{a}}{2H^2a^7}\bigg)\bigg).
\end{eqnarray}
\begin{figure} \centering
\epsfig{file=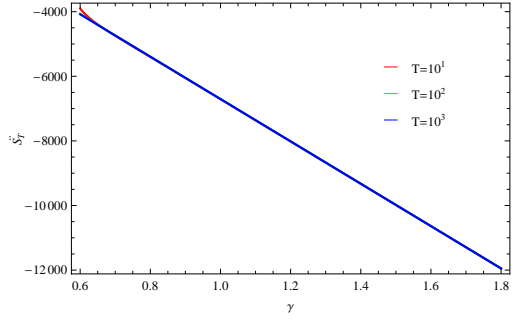,width=.50\linewidth}\caption{Plot of
$\ddot{S_T}$ versus $\gamma$ for logarithmic corrected entropy when
$\Gamma$ is a constant.}
\end{figure}The graphical behavior between $\ddot{S_T}$ and $\gamma$
is shown in Figure \textbf{6} for three values of $T$. We observe
that all the trajectories are showing the negative increasing
behavior, which exhibit the validity of thermodynamical
equilibrium.\\\\
\textbf{\underline{Power law corrected entropy:}}\\\\
To find the validity of thermodynamical equilibrium we take $\Gamma$
as constant for which second order differential equation can be
expressed with the help of Eq.(\ref{E29}) as
\begin{eqnarray}\label{E29a}
\ddot{S_T}&=&-\frac{\big(\frac{c^2}{a^6}+\gamma\big(1-\frac{\Gamma}{3H}\big)\big(3H^2-\frac{c^2}{2a^6}\big)\big)
\big(\frac{1}{4HL^2_p}-\frac{\big(2-\frac{\delta}{2}\big)\big(\frac{1}{H}\big)^{3-\delta}K_\delta}{4L^2_p}\big)\dot{H}}{H^3}\nonumber\\
&-&\bigg(\frac{-\gamma\big(1-\frac{\Gamma}{3H}\big)\big(3H^2-\frac{c^2}{2a^6}\big)+\frac{\gamma\big(3H^2-\frac{c^2}{2a^6}\big)
\big(\frac{c^2}{a^6}+\gamma\big(1-\frac{\Gamma}{3H}\big)\big(3H^2-\frac{c^2}{2a^6}\big)\big)}{2H^2}}{H^3T_0}\bigg)\nonumber\\
&\times&\dot{H}\big(H^2-\frac{c^2}{6a^6}\big)^{\frac{1-\gamma}{\gamma}}+(1-\gamma)\big(\frac{c^2\dot{a}}{a^7}+2H\dot{H}\big)
\big(H^2-\frac{c^2}{6a^6}\big)
^{-1+\frac{1-\gamma}{\gamma}}\nonumber\\
&\times&\bigg(\frac{-\gamma\big(1-\frac{\Gamma}{3H}\big)\big(3H^2-\frac{c^2}{2a^6}\big)+\frac{\gamma\big(3H^2-\frac{c^2}{2a^6}\big)
\big(\frac{c^2}{a^6}+\gamma\big(1-\frac{\Gamma}{3H}\big)\big(3H^2-\frac{c^2}{2a^6}\big)\big)}{2H^2}}{\gamma
H^2T_0}\bigg)\nonumber\\
&+&\frac{\bigg(\frac{c^2}{a^6}+\gamma\big(1-\frac{\Gamma}{3H}\big)\big(3H^2-\frac{c^2}{2a^6}\big)\bigg)
\bigg(-\frac{\dot{H}}{4HL^2_p}+\frac{(3-\delta)\dot{H}\big(2-\frac{\delta}{2}\big)\big(\frac{1}{H}\big)
^{4-\delta}K_\delta}{4L^2_p}\bigg)}{2H^2}\nonumber\\
&+&\bigg(\frac{1}{4HL^2_p}-\frac{\big(2-\frac{\delta}{2}\big)\big(\frac{1}{H}\big)^{3-\delta}K_\delta}{4HL^2_p}\bigg)
\bigg(\frac{\big(-\frac{6c^2\dot{a}}{a^7}\big)
+\gamma\big(1-\frac{\Gamma}{3H}\big)\big(\frac{3c^2\dot{a}}{a^7}+6H\dot{H}\big)}{2H^2}\nonumber\\
&+&\frac{\gamma\Gamma\dot{H}\big(3H^2-\frac{c^2}{2a^6}\big)}{6H^4}\bigg)+\frac{\big(H^2-\frac{c^2}{6a^6}\big)
^{-\frac{1-\gamma}{\gamma}}}{H^2T_0}\bigg(\frac{\gamma\Gamma\dot{H}\big(3H^2-\frac{c^2}{2a^6}\big)}{3H^2}
-\gamma\big(\frac{3c^2\dot{a}}{a^7}\nonumber\\
&+&6H\dot{H}\big)\big(1-\frac{\Gamma}{3H}\big)-\frac{\gamma\dot{H}\big(3H^2-\frac{c^2}{2a^6}\big)\big(\frac{c^2}{a^6}
+\gamma\big(1-\frac{\Gamma}{3H}\big)\big(3H^2-\frac{c^2}{2a^6}\big)\big)}{H^3}\nonumber\\
&+&\frac{\gamma\big(\frac{3c^2\dot{a}}{a^7}+6H\dot{H}\big)\big(\frac{c^2}{a^6}+\gamma\big(1-\frac{\Gamma}{3H}\big)
\big(3H^2-\frac{c^2}{2a^6}\big)\big)}{2H^2}+\gamma\big(3H^2-\frac{c^2}{2a^6}\big)\nonumber\\
&\times&\bigg(\frac{-\frac{6c^2\dot{a}}{a^7}+\frac{\gamma\Gamma\dot{H}\big(3H^2-\frac{c^2}{2a^6}\big)}{3H^2}
+\gamma\big(1-\frac{\Gamma}{3H}\big)\big(\frac{3c^2\dot{a}}{a^7}+6H\dot{H}\big)}{2H^2}\bigg)\bigg).
\end{eqnarray}
\begin{figure} \centering
\epsfig{file=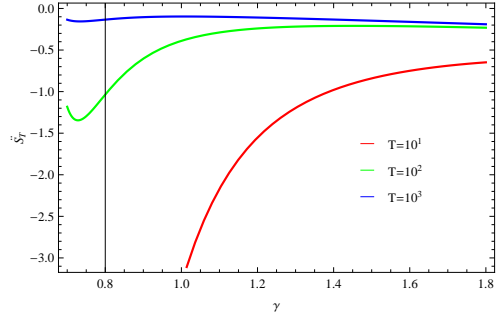,width=.50\linewidth}\caption{Plot of
$\ddot{S_T}$ versus $\gamma$ for power law corrected entropy when
$\Gamma$ is constant.}
\end{figure}The plot of $\ddot{S_T}$ versus $\gamma$ as shown in
Figure \textbf{7} for three values of $T$. We analyze that
thermodynamical equilibrium is satisfying the condition
$\ddot{S_T}<0$ for all values of $T$, which leads to validity of
thermodynamical equilibrium.\\\\
\textbf{\underline{Renyi entropy:}}\\\\
In present scenario, we observe the validity of thermodynamical
equilibrium by keeping $\Gamma$ as constant for which second order
differential equation is given by using Eq.(\ref{E38})
\begin{eqnarray}\label{E38a}
\ddot{S_T}&=&-\frac{16\dot{H}}{(\eta+8H^2)^2}\bigg(\frac{c^2}{a^6}+\gamma\big(3H^2-\frac{c^2}{2a^6}\big)
\big(1-\frac{\Gamma}{3H}\big)\bigg)-\frac{\dot{H}}{
H^2(\eta+8H^2)}\nonumber\\
&\times&\bigg(\frac{c^2}{a^6}+\gamma\big(3H^2-\frac{c^2}{2a^6}\big)
\big(1-\frac{\Gamma}{3H}\big)\bigg)-\frac{2\dot{H}\big(H^2-\frac{c^2}{6a^6}\big)^{\frac{1-\gamma}{\gamma}}}{H^3T_0}
\bigg(\big(3H^2-\frac{c^2}{2a^6}\big)\nonumber\\
&\times&\big(1-\frac{\Gamma}{3H}\big)(-\gamma)+\frac{\gamma\big(3H^2-\frac{c^2}{2a^6}\big)
\big(\frac{c^2}{a^6}+\gamma\big(3H^2-\frac{c^2}{2a^6}\big)\big(1-\frac{\Gamma}{3H}\big)\big)}{2H^2}\bigg)\nonumber\\
&+&\bigg(\frac{\gamma\big(3H^2-\frac{c^2}{2a^6}\big)
\big((\frac{c^2}{a^6}+\gamma\big(3H^2-\frac{c^2}{2a^6}\big)\big(1-\frac{\Gamma}{3H}\big)\big)}{2H^2}
-\gamma\big(3H^2-\frac{c^2}{2a^6}\big)\nonumber\\
&\times&\big(1-\frac{\Gamma}{3H}\big)\bigg)\frac{(1-\gamma)\big(\frac{c^2\dot{a}}{a^7}+2H\dot{H}\big)
\big(H^2-\frac{c^2}{6a^6}\big)^{-1+\frac{1-\gamma}{\gamma}}}{\gamma
H^2T_0}+\frac{1}{H(\eta+8H^2)}\nonumber\\
&\times&\bigg(-\frac{6c^2\dot{a}}{a^7}+\frac{\gamma\Gamma\dot{H}\big(3H^2-\frac{c^2}{2a^6}\big)}{3H^2}
+\gamma\big(1-\frac{\Gamma}{3H}\big)\big(\frac{3c^2\dot{a}}{a^7}+6H\dot{H}\big)\bigg)+\frac{1}{H^2T_0}\nonumber\\
&\times&\big(H^2-\frac{c^2}{6a^6}\big)^{\frac{1-\gamma}{\gamma}}\bigg(-\frac{\gamma\dot{H}
\big(3H^2-\frac{c^2}{2a^6}\big)\big(\frac{c^2}{a^6}+\gamma\big(3H^2-\frac{c^2}{2a^6}\big)
\big(1-\frac{\Gamma}{3H}\big)\big)}{H^3}\nonumber\\
&+&\frac{\gamma\big(\frac{3c^2\dot{a}}{a^7}+6H\dot{H}\big)
\big(\frac{c^2}{a^6}+\gamma\big(3H^2-\frac{c^2}{2a^6}\big)\big(1-\frac{\Gamma}{3H}\big)\big)}{2H^2}
-\frac{\gamma\Gamma\dot{H}\big(3H^2-\frac{c^2}{2a^6}\big)}{3H^2}\nonumber\\
&-&\gamma\big(1-\frac{\Gamma}{3H}\big)\big(\frac{3c^2\dot{a}}{a^7}+6H\dot{H}\big)
+\bigg(-\frac{6c^2\dot{a}}{a^7}+\frac{\gamma\Gamma\dot{H}\big(3H^2-\frac{c^2}{2a^6}\big)}{3H^2}
+\big(1-\frac{\Gamma}{3H}\big)\nonumber\\
&\times&\gamma\big(\frac{3c^2\dot{a}}{a^7}+6H\dot{H}\big)\bigg)\frac{\gamma\big(3H^2-\frac{c^2}{2a^6}\big)}{2H^2}\bigg).
\end{eqnarray}
\begin{figure} \centering
\epsfig{file=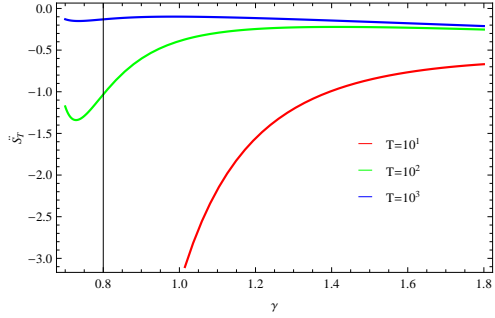,width=.50\linewidth}\caption{Plot of
$\ddot{S_T}$ versus $\gamma$ for Renyi entropy when
$\Gamma$=constant.}
\end{figure}The graphical behavior of $\ddot{S_T}$ versus $\gamma$ for three
values of $T$ as shown in Figure \textbf{8}. It can be seen that
$\ddot{S_T}$ is negative for all values of $T$ satisfying the
condition $d^2{S_T}/dt^2<0$ which exhibits the thermodynamical
equilibrium.

\subsection{$\Gamma=\Gamma(t)$}
Secondly, we consider particle creation rate $\Gamma$ as a variable,
i.e., $\Gamma=\Gamma(t)$.\\\\
\textbf{\underline{Bekenstein entropy:}}\\\\
The differentiation of Eq.(\ref{E21}) leads to
\begin{eqnarray}\label{E21b}
\ddot{S_T}&=&-\frac{3\dot{H}\big(\frac{c^2}{a^6}+\gamma\big(1-\frac{\Gamma}{3H}\big)\big(-\frac{c^2}{2a^6}+
3H^2\big)\big)}{8H^4}-2\dot{H}\big(-\frac{c^2}{6a^6}+H^2\big)^{\frac{1-\gamma}{\gamma}}\nonumber\\
&\times&\bigg(\frac{-\gamma\big(3H^2-\frac{c^2}{2a^6}\big)\big(1-\frac{\Gamma}{3H}\big)+\frac{\gamma\big(3H^2-\frac{c^2}{2a^6}\big)
\big(\frac{c^2}{a^6}+\gamma\big(3H^2-\frac{c^2}{2a^6}\big)\big(1-\frac{\Gamma}{3H}\big)\big)}{2H^2}}{H^3T_0}\bigg)\nonumber\\
&+&\bigg(\frac{-\gamma\big(3H^2-\frac{c^2}{2a^6}\big)\big(1-\frac{\Gamma}{3H}\big)+\frac{\gamma\big(3H^2-\frac{c^2}{2a^6}\big)
\big(\frac{c^2}{a^6}+\gamma\big(3H^2-\frac{c^2}{2a^6}\big)\big(1-\frac{\Gamma}{3H}\big)\big)}{2H^2}}{\gamma
H^2T_0}\bigg)\nonumber\\
&\times&(1-\gamma)\big(\frac{c^2\dot{a}}{a^7}+2H\dot{H}\big)\big(H^2-\frac{c^2}{6a^6}\big)^{-1+\frac{1-\gamma}{\gamma}}
-\frac{6c^2\dot{a}}{8H^3a^7}+\frac{\big(\frac{3c^2\dot{a}}{a^7}+6H\dot{H}\big)}{8H^3}\nonumber\\
&\times&\gamma\big(1-\frac{\Gamma}{3H}\big)+\frac{\gamma\big(3H^2-\frac{c^2}{2a^6}\big)\big(\frac{\Gamma\dot{H}}{3H^2}
-\frac{\dot{\Gamma}}{3H}\big)}{8H^3}+\frac{\big(H^2-\frac{c^2}{6a^6}\big)^{\frac{1-\gamma}{\gamma}}}{H^2T_0}\nonumber\\
&\times&\bigg(-\frac{\gamma\dot{H}\big(3H^2-\frac{c^2}{2a^6}\big)\big(\frac{c^2}{a^6}+\big(3H^2-\frac{c^2}{2a^6}\big)
\big(1-\frac{\Gamma}{3H}\big)\big)}{H^3}-\gamma\big(1-\frac{\Gamma}{3H}\big)\nonumber\\
&\times&\big(\frac{3c^2\dot{a}}{a^7}+6H\dot{H}\big)+\frac{\gamma\big(\frac{c^2}{a^6}+\gamma\big(3H^2-\frac{c^2}{2a^6}\big)
\big(1-\frac{\Gamma}{3H}\big)\big)\big(\frac{3c^2\dot{a}}{a^7}+6H\dot{H}\big)}{2H^2}\nonumber\\
&-&\gamma\big(3H^2-\frac{c^2}{2a^6}\big)\big(\frac{\Gamma\dot{H}}{3H^2}-\frac{\dot{\Gamma}}{3H}\big)
+\bigg(\frac{-\frac{6c^2\dot{a}}{a^7}+\gamma\big(1-\frac{\Gamma}{3H}\big)\big(\frac{3c^2\dot{a}}{a^7}+6H\dot{H}\big)}{2H^2}\nonumber\\
&+&\frac{\gamma\big(3H^2-\frac{c^2}{2a^6}\big)\big(\frac{\Gamma\dot{H}}{3H^2}-\frac{\dot{\Gamma}}{3H}\big)}{2H^2}
\bigg)\big(3H^2-\frac{c^2}{2a^6}\big)\gamma\bigg).
\end{eqnarray}
\begin{figure} \centering
\epsfig{file=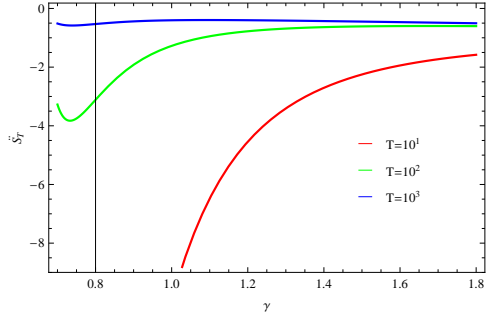,width=.50\linewidth}\caption{Plot of
$\ddot{S_T}$ versus $\gamma$ for Bekenstein entropy when
$\Gamma=\Gamma (t)$.}
\end{figure}The plot between $\ddot{S_T}$ versus $\gamma$ for three values of
$T$ as shown in Figure \textbf{9} by keeping the constant values
same as for constant $\Gamma$. It is observed that the
thermodynamical is obeying the condition $\ddot{S_T}<0$ which leads
to the validity of thermodynamical equilibrium.\\\\
\textbf{\underline{Logarithmic corrected entropy:}}\\\\
We discuss the stability analysis of thermal equilibrium in the
presence of logarithmic corrected entropy by taking $\Gamma$ as
variable for which Eq.(\ref{E25}) reduces to second order
differential equation as
\begin{eqnarray}\label{E25b}
\ddot{S_T}&=&-\frac{\dot{H}\big(\frac{1}{4HL^2_p}+2\alpha
H-16L^2_p\beta
H^3\big)\big(\frac{c^2}{a^6}+\gamma\big(1-\frac{\Gamma}{3H}\big)\big(3H^2-\frac{c^2}{2a^6}\big)\big)}{H^3}\nonumber\\
&-&\bigg(\frac{-\gamma\big(1-\frac{\Gamma}{3H}\big)
\big(3H^2-\frac{c^2}{2a^6}\big)+\frac{\gamma\big(3H^2-\frac{c^2}{2a^6}\big)\big(\frac{c^2}{a^6}
+\gamma\big(1-\frac{\Gamma}{3H}\big)\big(3H^2-\frac{c^2}{2a^6}\big)\big)}{2H^2}}{H^3T_0}\bigg)\nonumber\\
&\times&2\dot{H}\big(H^2-\frac{c^2}{6a^6}\big)^{\frac{1-\gamma}{\gamma}}+(1-\gamma)\big(\frac{c^2\dot{a}}{a^7}+2H\dot{H}\big)
\big(H^2-\frac{c^2}{6a^6}\big)
^{-1+\frac{1-\gamma}{\gamma}}\nonumber\\
&\times&\bigg(\frac{-\gamma\big(1-\frac{\Gamma}{3H}\big)
\big(3H^2-\frac{c^2}{2a^6}\big)+\frac{\gamma\big(3H^2-\frac{c^2}{2a^6}\big)\big(\frac{c^2}{a^6}
+\gamma\big(1-\frac{\Gamma}{3H}\big)\big(3H^2-\frac{c^2}{2a^6}\big)\big)}{2H^2}}{\gamma
H^2T_0}\bigg)\nonumber\\
&+&\frac{\big(\frac{c^2}{a^6}+\gamma\big(1-\frac{\Gamma}{3H}\big)\big(3H^2-\frac{c^2}{2a^6}\big)\big)
\big(2\alpha\dot{H}-\frac{\dot{H}}{4L^2_p H^2}-48L^2_p\beta
H^2\dot{H}\big)}{2H^2}\nonumber\\
&+&\big(\frac{1}{4HL^2_p}+2\alpha H-16L^2_p\beta
H^3\big)\bigg(\frac{-\frac{6c^2\dot{a}}{a^7}+
\gamma\big(1-\frac{\Gamma}{3H}\big)\big(\frac{3c^2\dot{a}}{a^7}+6H\dot{H}\big)}{2H^2}\nonumber\\
&+&\frac{\gamma\big(3H^2-\frac{c^2}{2a^6}\big)\big(\frac{\Gamma\dot{H}}{3H^2}-\frac{\dot{\Gamma}}{3H}\big)}{2H^2}\bigg)
+\frac{1}{H^2T_0}\big(H^2-\frac{c^2}{6a^6}\big)
^{\frac{1-\gamma}{\gamma}}\bigg(-\big(3H^2-\frac{c^2}{2a^6}\big)\nonumber\\
&\times&\frac{\gamma\dot{H}\big(\frac{c^2}{a^6}+\gamma\big(1-\frac{\Gamma}{3H}\big)
\big(3H^2-\frac{c^2}{2a^6}\big)\big)}{H^3}-\gamma\big(1-\frac{\Gamma}{3H}\big)\big(\frac{3c^2\dot{a}}{a^7}+6H\dot{H}\big)\nonumber\\
&+&\frac{\gamma\big(\frac{c^2}{a^6}+\gamma\big(1-\frac{\Gamma}{3H}\big)\big(3H^2-\frac{c^2}{2a^6}\big)\big)
\big(\frac{3c^2\dot{a}}{a^7}+6H\dot{H}\big)}{2H^2}-\gamma\big(3H^2-\frac{c^2}{2a^6}\big)\nonumber\\
&\times&\big(\frac{\Gamma\dot{H}}{3H^2}-\frac{\dot{\Gamma}}{3H}\big)+\frac{\gamma\big(3H^2-\frac{c^2}{2a^6}\big)
\big(-\frac{6c^2\dot{a}}{a^7}+\gamma\big(1-\frac{\Gamma}{3H}\big)\big(\frac{3c^2\dot{a}}{a^7}+6H\dot{H}\big)
\big)}{2H^2}\nonumber\\
&+&\frac{\gamma^2\big(3H^2-\frac{c^2}{2a^6}\big)^2\big(\frac{\Gamma\dot{H}}{3H^2}-\frac{\dot{\Gamma}}{3H}\big)}{2H^2}\bigg).
\end{eqnarray}
\begin{figure} \centering
\epsfig{file=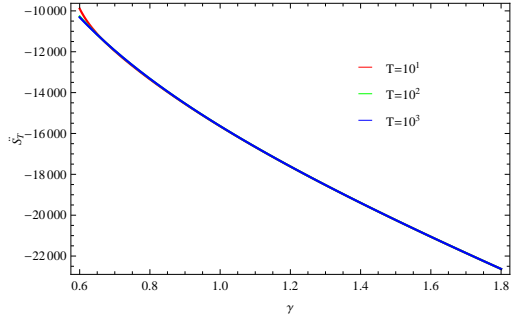,width=.50\linewidth}\caption{Plot of
$\ddot{S_T}$ versus $\gamma$ when logarithmic corrected entropy
$\Gamma=\Gamma (t)$.}
\end{figure}The plot of $\ddot{S_T}$ versus $\gamma$ as shown in Figure
\textbf{10} for three values of $T$ by keeping the same values as in
above case, we observe that thermal equilibrium condition
$\ddot{S_T}<0$ fulfill which leads to the validity of
thermodynamical equilibrium.\\\\
\textbf{\underline{Power law corrected entropy:}}\\\\
For variable $\Gamma$, the differentiation of Eq.(\ref{E29}) turns
out to be
\begin{eqnarray}\label{E29b}
\ddot{S_T}&=&-\frac{\big(\frac{1}{4HL^2_p}-\frac{\big(2-\frac{\delta}{2}\big)\big(\frac{1}{H}\big)^{3-\delta}K_\delta}{4L^2_p}\big)
\big(\frac{c^2}{a^6}+\gamma\big(1-\frac{\Gamma}{3H}\big)\big(3H^2-\frac{c^2}{2a^6}\big)\big)\dot{H}}{H^3}\nonumber\\
&-&\bigg(\frac{-\gamma\big(1-\frac{\Gamma}{3H}\big)\big(3H^2-\frac{c^2}{2a^6}\big)+\frac{\gamma\big(3H^2-\frac{c^2}{2a^6}\big)
\big(\frac{c^2}{a^6}+\gamma\big(1-\frac{\Gamma}{3H}\big)\big(3H^2-\frac{c^2}{2a^6}\big)\big)}{2H^2}}{H^3T_0}\bigg)\nonumber\\
&\times&2\dot{H}\big(H^2-\frac{c^2}{6a^6}\big)^{\frac{1-\gamma}{\gamma}}+(1-\gamma)\big(\frac{c^2\dot{a}}{a^7}+2H\dot{H}\big)
\big(H^2-\frac{c^2}{6a^6}\big)^{-1+\frac{1-\gamma}{\gamma}}\nonumber\\
&\times&\bigg(\frac{-\gamma\big(3H^2-\frac{c^2}{2a^6}\big)\big(1-\frac{\Gamma}{3H}\big)+\frac{\gamma\big(3H^2-\frac{c^2}{2a^6}\big)
\big(\frac{c^2}{a^6}+\gamma\big(3H^2-\frac{c^2}{2a^6}\big)\big(1-\frac{\Gamma}{3H}\big)\big)}{2H^2}}{\gamma
H^2T_0}\bigg)\nonumber\\
&+&\frac{\big(-\frac{\dot{H}}{4HL^2_p}+\frac{\dot{H}(3-\delta)\big(2-\frac{\delta}{2}\big)
\big(\frac{1}{H}\big)^{4-\delta}K_\delta}{4L^2_p}\big)\bigg(\frac{c^2}{a^6}+\gamma\big(3H^2-\frac{c^2}{2a^6}\big)
\big(1-\frac{\Gamma}{3H}\big)\bigg)}{2H^2}\nonumber\\
&+&\big(\frac{1}{4HL^2_p}-\frac{\big(2-\frac{\delta}{2}\big)\big(\frac{1}{H}\big)^{3-\delta}K_\delta}{4L^2_p}\big)
\bigg(\frac{-\frac{6c^2\dot{a}}{a^7}+\gamma\big(1-\frac{\Gamma}{3H}\big)\big(\frac{3c^2\dot{a}}{a^7}+6H\dot{H}\big)}{2H^2}\nonumber\\
&+&\frac{\gamma\big(3H^2-\frac{c^2}{2a^6}\big)\big(\frac{\Gamma\dot{H}}{3H^2}-\frac{\dot{\Gamma}}{3H}\big)}{2H^2}\bigg)
+\frac{\big(H^2-\frac{c^2}{6a^6}\big)^{\frac{1-\gamma}{\gamma}}}{H^2T_0}\bigg(-\big(\frac{3c^2\dot{a}}{a^7}+6H\dot{H}\big)\nonumber\\
&\times&\gamma\big(1-\frac{\Gamma}{3H}\big)-\frac{\gamma\dot{H}\big(3H^2-\frac{c^2}{2a^6}\big)
\big(\frac{c^2}{a^6}+\gamma\big(3H^2-\frac{c^2}{2a^6}\big)\big(1-\frac{\Gamma}{3H}\big)\big)}{H^3}\nonumber\\
&+&\frac{\gamma\big(\frac{3c^2\dot{a}}{a^7}+6H\dot{H}\big)\big(\frac{c^2}{a^6}+\gamma\big(3H^2-\frac{c^2}{2a^6}\big)
\big(1-\frac{\Gamma}{3H}\big)\big)}{2H^2}-\gamma\big(3H^2-\frac{c^2}{2a^6}\big)\nonumber\\
&\times&\big(\frac{\Gamma\dot{H}}{3H^2}-\frac{\dot{\Gamma}}{3H}\big)+\frac{\gamma\big(3H^2-\frac{c^2}{2a^6}\big)}{2H^2}
\bigg(+\gamma\big(1-\frac{\Gamma}{3H}\big)\big(\frac{3c^2\dot{a}}{a^7}+6H\dot{H}\big)\nonumber\\
&+&\gamma\big(3H^2-\frac{c^2}{2a^6}\big)\big(\frac{\Gamma\dot{H}}{3H^2}-\frac{\dot{\Gamma}}{3H}\big)-\frac{6c^2\dot{a}}{a^7}\bigg)\bigg).
\end{eqnarray}
\begin{figure} \centering
\epsfig{file=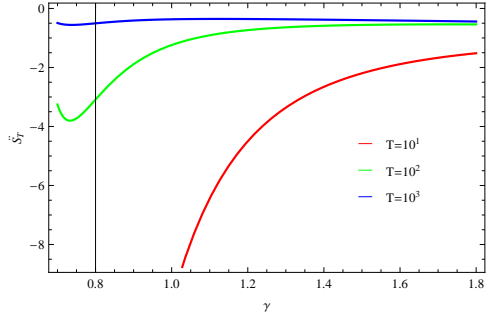,width=.50\linewidth}\caption{Plot of
$\ddot{S_T}$ versus $\gamma$ for power law correction when
$\Gamma=\Gamma (t)$}
\end{figure}The plot of $\ddot{S_T}$ versus $\gamma$ for three values of $T$ by keeping the
same values as above mentioned (Figure \textbf{11}), one can observe
easily the validity of thermodynamical equilibrium for all values of
$T$ with $\ddot{S_T}<0$.\\\\
\textbf{\underline{Renyi entropy:}}\\\\
For variable $\Gamma$, Eq.(\ref{E38}) gives
\begin{eqnarray}\label{E38b}
\ddot{S_T}&=&-\frac{16\dot{H}}{(\eta+8H^2)^2}\bigg(\frac{c^2}{a^6}+\gamma\big(3H^2-\frac{c^2}{2a^6}\big)
\big(1-\frac{\Gamma}{3H}\big)\bigg)-\frac{\dot{H}}{
H^2(\eta+8H^2)}\nonumber\\
&\times&\bigg(\frac{c^2}{a^6}+\gamma\big(3H^2-\frac{c^2}{2a^6}\big)
\big(1-\frac{\Gamma}{3H}\big)\bigg)-\frac{2\dot{H}\big(H^2-\frac{c^2}{6a^6}\big)^{\frac{1-\gamma}{\gamma}}}{H^3T_0}
\bigg(\big(3H^2-\frac{c^2}{2a^6}\big)\nonumber\\
&\times&\big(1-\frac{\Gamma}{3H}\big)(-\gamma)+\frac{\gamma\big(3H^2-\frac{c^2}{2a^6}\big)
\big(\frac{c^2}{a^6}+\gamma\big(3H^2-\frac{c^2}{2a^6}\big)\big(1-\frac{\Gamma}{3H}\big)\big)}{2H^2}\bigg)\nonumber\\
&+&\bigg(\frac{\gamma\big(3H^2-\frac{c^2}{2a^6}\big)
\big(\frac{c^2}{a^6}+\gamma\big(3H^2-\frac{c^2}{2a^6}\big)\big(1-\frac{\Gamma}{3H}\big)\big)}{2H^2}
-\gamma\big(3H^2-\frac{c^2}{2a^6}\big)\nonumber\\
&\times&\big(1-\frac{\Gamma}{3H}\big)\bigg)\frac{(1-\gamma)\big(\frac{c^2\dot{a}}{a^7}+2H\dot{H}\big)
\big(H^2-\frac{c^2}{6a^6}\big)^{-1+\frac{1-\gamma}{\gamma}}}{\gamma
H^2T_0}+\frac{1}{H(\eta+8H^2)}\nonumber\\
&\times&\bigg(\gamma(1-\frac{\Gamma}{3H})(\frac{3c^2\dot{a}}{a^7}+6H\dot{H})
+\gamma(3H^2-\frac{c^2}{2a^6})(\frac{\Gamma\dot{H}}{3H^2}-\frac{\dot{\Gamma}}{3H})-\frac{6c^2\dot{a}}{a^7}\bigg)\nonumber\\
&+&\frac{\big(H^2-\frac{c^2}{6a^6}\big)^{\frac{1-\gamma}{\gamma}}}{H^2T_0}\bigg(-\frac{\gamma\dot{H}
\big(3H^2-\frac{c^2}{2a^6}\big)\big(\frac{c^2}{a^6}+\gamma\big(3H^2-\frac{c^2}{2a^6}\big)\big(1-\frac{\Gamma}{3H}\big)\big)}{H^3}\nonumber\\
&+&\frac{\gamma\big(\frac{3c^2\dot{a}}{a^7}+6H\dot{H}\big)
\big(\frac{c^2}{a^6}+\gamma\big(3H^2-\frac{c^2}{2a^6}\big)\big(1-\frac{\Gamma}{3H}\big)\big)}{2H^2}
-\gamma\big(\frac{3c^2\dot{a}}{a^7}+6H\dot{H}\big)\nonumber\\
&\times&\big(1-\frac{\Gamma}{3H}\big)+\frac{\gamma\big(3H^2-\frac{c^2}{2a^6}\big)}{2H^2}
\bigg(-\frac{6c^2\dot{a}}{a^7}+\gamma\big(1-\frac{\Gamma}{3H}\big)\big(\frac{3c^2\dot{a}}{a^7}+6H\dot{H}\big)\nonumber\\
&+&\gamma\big(3H^2-\frac{c^2}{2a^6}\big)\big(\frac{\Gamma\dot{H}}{3H^2}-\frac{\dot{\Gamma}}{3H}\big)\bigg)
-\gamma\big(3H^2-\frac{c^2}{2a^6}\big)\big(\frac{\Gamma\dot{H}}{3H^2}-\frac{\dot{\Gamma}}{3H}\big)\bigg).
\end{eqnarray}
\begin{figure} \centering
\epsfig{file=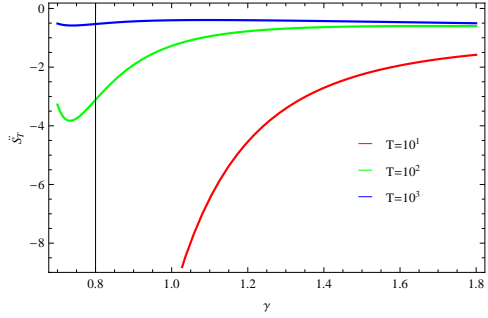,width=.50\linewidth}\caption{Plot of
$\ddot{S_T}$ versus $\gamma$ for Renyi entropy when $\Gamma=\Gamma
(t)$}
\end{figure}The graphical behavior of $\ddot{S_T}$ versus $\gamma$ for three
values of $T$ as shown in Figure \textbf{12} by keeping the constant
values same as in above. One can see that $\ddot{S_T}$ is negative
for all values of $T$ satisfying the condition $d^2{S_T}<0$ which
leads to the validity of thermodynamical equilibrium.

\section{Comparison}

Here we provide some literature on underlying scenario for
comparison and summary of present work.

Harko et al. \cite{harko} considered the possibility of a
gravitationally induced particle production through the mechanism of
a non-minimal curvature–matter coupling. Firstly, they have
reformulated the model in terms of an equivalent scalar–tensor
theory, with two arbitrary potentials. The particle number creation
rates, the creation pressure, and the entropy production rates have
explicitly obtained as functions of the scalar field and its
potentials as well as of the matter Lagrangian. The temperature
evolution laws of the newly created particles have also obtained.
The cosmological implications of the model have briefly investigated
and it is shown that the late-time cosmic acceleration may be due to
particle creation process. Furthermore, it has also shown that due
to the curvature–matter coupling, during the cosmological evolution
a large amount of comoving entropy is also produced.

Mitra et al. \cite{R3} have studied thermodynamics laws by assuming
flat FRW universe enveloped by by apparent and event horizon in the
framework of RSII brane model and DGP brane scenario. Assuming
extended Hawking temperature on the horizon, the unified first law
is examined for perfect fluid (with constant equation of state) and
Modified Chaplygin Gas model. As a result there is a modification of
Bekenstein entropy on the horizons. Further the validity of the GSLT
and thermodynamical equilibrium have also been investigated. By
assuming the gravitationally induced particle scenario with constant
specific entropy and arbitrary particle creation rate ($\Gamma$),
thermodynamics on the apparent horizon for FRW universe has been
discussed \cite{1j}. They have investigated the first law, GSLT and
thermodynamical equilibrium by assuming the EoS for perfect fluid
and put forward various constraints on $\Gamma$ for which
thermodynamical laws hold.

Recently, we have done the thermodynamical analysis for
gravitationally induced particle creation scenario in the framework
of DGP braneworld model \cite{js} by assuming usual entropy as well
as its entropy corrections (power law and logarithmic) in the flat
FRW universe. We have extracted EoS parameter and obtained its
various constraints with respect to quintessence, vacuum and phantom
era of the universe. For variable as well as constant particle
creation rate ($\Gamma$), the first law of thermodynamics, GSLT and
thermal equilibrium condition is satisfied in all the cases of
entropy forms within some specific ranges of $\gamma$. In the
present work, we have extended this work in the dynamical
Chern-Simons modified gravity taking equation of state for perfect
fluid as $p=(\gamma-1)\rho$. By assuming various modified entropies
(Bekenstein, logarithmic, power law correction and Reyni), we
investigate the first law of thermodynamics, equilibrium condition
and generalized second law of thermodynamics on the apparent horizon
in the presence of particle creation rate. It is concluded that the
GSLT and thermodynamical equilibrium are satisfying the conditions
$\frac{dS_T}{dt}\geq0$ and $\frac{d^2S_T}{dt^2}<0$ for all values of
$T$ throughout the range $\frac{2}{3}\leq\gamma\leq 2$.

In section 4, we have discussed the thermal equilibrium phenomenon
for $\Gamma$ as variable and constant. The cosmic history is
well-established through different observational sources that the
radiation phase was followed by a matter dominated era which
eventually passed through to a second de Sitter phase. Accordingly,
it can be expected that in the radiation dominated era the entropy
increased but the thermodynamic equilibrium was not achieved
\cite{31++}. If this were not true, the universe would have attained
a state of maximum entropy and would have stayed in it forever
unless acted upon by some "external agent." However, it is a
well-known fact \cite{dd} that the production of particles was
suppressed during the radiation phase, so in this model there would
be no external agent to remove the system from thermodynamic
equilibrium. In the present work, it is very difficult to find the
analytical constraints to meet the equilibrium condition as
discussed in the \cite{saha} due to lengthy expressions of
$\ddot{S_T}$. Therefore, we checked these conditions graphically
taking specific values of model parameters.

Since the prefect fluid is the simplest model in the cosmological
studies, we study the prefect fluid case. In fact, this model can
provide suitable results in the Einstein theory as well as modified
theories \cite{grg,plb}. Moreover, the motivation of the present
work in the framework of the particle creation mechanism comes from
some recent related works. It has been shown in \cite{391, 401} that
the entire cosmic evolution from inflationary stage can be described
by particle creation mechanism with some specific choices of the
particle creation rates. As these works show late-time acceleration
without any concept of dark energy, so, it is very interesting to
think of the particle creation mechanism as an alternative way of
explaining the idea of dark energy.

\section{Conclusion}

In this work, we have investigated the validity of first law of
thermodynamics, GSLT and thermodynamical equilibrium for particle
creation scenario in the presence of perfect fluid EoS
$p=(\gamma-1)\rho$ by assuming the different entropy corrections
such as Bakenstein entropy, logarithmic corrected entropy and power
law corrected entropy and Renyi entropy in a newly proposed
dynamical Chern-Simons modified gravity. We have summarized our
results as follows:
\begin{itemize}
\item \textbf{For Bekenstein entropy}\\
We have analyzed that first law of thermodynamics is showing the
validity for $\Gamma=3H\bigg(\frac{c^2}{\gamma
a^6}\big(3H^2-\frac{c^2}{2a^6}\big)^{-1}\bigg)$. However, GSLT
remains valid for all values of $T$ with $\frac{2}{3}\leq\gamma\leq
2$. Further, we have analyzed the validity of thermodynamical
equilibrium for constant and variable $\Gamma$. From
Figures(\textbf{5} and \textbf{9}), we observe that thermodynamical
equilibrium is satisfying the condition $\frac{d^2S_T}{dt^2}<0$ for
all values of $T$ with $\frac{2}{3}\leq\gamma\leq 2$.
\item \textbf{For Logarithmic corrected Entropy}\\
In the presence of logarithmic  corrected entropy it can be seen
that the first law of thermodynamics is valid on the apparent
horizon when
$\Gamma=3H\bigg(1-\frac{1}{\gamma}\big(3H^2-\frac{c^2}{a^6}\big)^{-1}
\bigg(\frac{\gamma\big(3H^2-\frac{c^2}{2a^6}\big)}{\big(\frac{1}{L^2_{\textmd{p}}}+8H^2\alpha-64\beta
H^4L^2_{\textmd{p}}\big)}-\frac{c^2}{a^6}\bigg)\bigg)$. We have also
investigated the validity of GSLT on apparent horizon satisfying the
condition $\frac{dS_T}{dt}\geq0$ (Figure \textbf{2}). The graphical
behavior of $\ddot{S_T}$ versus $\gamma$ as shown in Figures
(\textbf{6} and \textbf{10}). We observe the validity of
thermodynamical equilibrium for all values of $T$ for all values of
$\gamma$ for constant as well as variable $\Gamma$.
\item \textbf{Power law corrected entropy}\\
For power law corrected entropy we have investigates that first law
of thermodynamics is hold at apparent horizon for
$\Gamma=3H\bigg(1-\frac{1}{\gamma}\big(3H^2-\frac{c^2}{2a^6}\big)^{-1}\big(\gamma\big(3H^2-\frac{c^2}{2a^6}\big)\big)
\big(\frac{1}{L^2_{\textmd{p}}}-\big(2-\frac{\delta}{2}\big)\frac{K_\delta}{L^2_p}
\big(\frac{1}{H}\big)^{2-\delta}\big)^{-1}-\frac{c^2}{a^6}\bigg).$
From Figure \textbf{3} we can analyzed that the GSLT is valid for
all values of $T$ with $\frac{2}{3}\leq\gamma\leq 2$. Further, we
have investigated the validity of thermodynamical equilibrium
obeying the condition $\ddot{S_T}<0$ as shown in Figures(\textbf{7}
and \textbf{11}) for all values of $T$ with
$\frac{2}{3}\leq\gamma\leq 2$ for both constant and variable
$\Gamma$.
\item \textbf{For Renyi Entropy}\\
In this entropy we have observed that first law of thermodynamics is
holds when
$\Gamma=3H\big(1-\frac{1}{\gamma}\big(3H^2-\frac{c^2}{2a^6}\big)^{-1}\big(\frac{\gamma(\eta+8H^2)}{8H^2}
(3H^2-\frac{c^2}{2a^6})-\frac{c^2}{a^6}\big)\big)$. The Graphical
behavior of Figure \textbf{12} shows that all trajectories remains
positive for all values of $T$ with $\frac{2}{3}\leq\gamma\leq 2$
which leads to the validity of GSLT. Moreover, thermodynamical
equilibrium condition satisfied for all values of $T$ with all
values of $\gamma$ for constant and variable $\Gamma$.
\end{itemize}

\end{document}